\documentclass[manuscript]{aastex6}

\usepackage{graphicx}
\usepackage{natbib}

\newcommand{\ahat}{586.88}
\newcommand{\bhat}{0.91}

\begin{document}

\title{The Non-homogeneous Poisson Process for Fast Radio Burst Rates}
\author{Earl Lawrence, Scott Vander Wiel} 
\affil{Statistical Sciences \\ Los Alamos National Laboratory \\ P.O. Box 1663, MS F600 \\ Los Alamos NM 87545 \\ earl@lanl.gov}

\author{Casey Law} 
\affil{Department of Astronomy and Radio Astronomy Lab \\ University of California \\ Berkeley CA}

\author{Sarah Burke Spolaor}
\affil{National Radio Astronomy Observatory \\ NRAO Array Operation Center \\ P.O.Box O, 1003 Lopezville Road \\  Socorro NM 87801}

\and

\author{Geoffrey C. Bower}
\affil{Academia Sinica Institute of Astronomy and Astrophysics\\
645 N. A'ohoku Place\\
Hilo, HI 96720, USA}


\begin{abstract}
This paper presents the non-homogeneous Poisson process (NHPP) for modeling the rate of fast radio bursts (FRBs) and other infrequently observed astronomical events. The NHPP, well-known in statistics, can model changes in the rate as a function of both astronomical features and the details of an observing campaign. This is particularly helpful for rare events like FRBs because the NHPP can combine information across surveys, making the most of all available information. The goal of the paper is two-fold. First, it is intended to be a tutorial on the use of the NHPP. Second, we build an NHPP model that incorporates beam patterns and a power law flux distribution for the rate of FRBs. Using information from 12 surveys including 15 detections, we find an all-sky FRB rate of $\ahat$ events per sky per day above a flux of 1 Jy (95\% CI: 271.86, 923.72) and a flux power-law index of $\bhat$ (95\% CI: 0.57, 1.25). Our rate is lower than other published rates, but consistent with the rate given in \cite{champion2016five}.
\end{abstract}

\section{Introduction} \label{intro}
Rates for astronomical events are typically computed using only some of the available information. When some part of the observed parameter space (e.g., sensitivity) is non-ideal, astronomers cut that space out prior to estimating a rate (e.g., ``completeness limit''). This simplifies the analysis, but is dangerous when events are observed below the conservative threshold. Many compelling and active fields of astronomical research are subject to strong observational bias and low numbers of events, such as exoplanets \citep{foreman2014exoplanet}, gravitational waves \citep{abbott2016observation}, and near-earth asteroids \citep{stuart2004bias}.

The study of fast radio bursts (FRBs) is another good example. See for example \cite{lorimer2007bright}, \cite{thornton2013population}, or \cite{burke2014galactic}. To date, 17 FRBs, discovered in millisecond radio transient surveys, are listed in the FRB Catalog \footnote{\url{http://www.astronomy.swin.edu.au/pulsar/frbcat/}} of \cite{petroff2016frbcat}. In many of the detection papers, the rate is estimated by scaling the number of observed FRB detections by the product of the instrument observing area and the time-on-sky. This rate is generally assumed to be a cumulative rate for all events above a certain flux chosen to be the minimum detectable flux at the FWHM sensitivity for the instrument. However, the actual flux values contain information about the how rate changes as a function of flux.

This approach, correct but incomplete, can be limiting. This is especially true for FRB rate calculations because few have been observed thus far and we want to use all available information in the rate computations. For example, we know that radio telescopes have a sensitivity pattern (the ``beam'') that drops with distance, so there is more observing area at lower sensitivity and we are more likely to observe fluxes that are smaller. Furthermore, since most of the Universe is far away, we expect to see a much higher number of faint FRBs. The first effect reduces the flux that we record, while the second reduces the actual flux that reaches Earth. Both effects lead us to expect more events at one flux than we would at a higher flux.

The paper presents the use of nonhomogeneous Poisson process (NHPP) for computing the rate of fast radio bursts. The approach, frequently used in statistical modeling, will address the issues raised above and could be extended to address other issues that ultimately cause the probability of detection to vary. The method is applicable to event searches of any kind. The paper has two goals. One, it is intended to be a tutorial in the use of nonhomogeneous Poisson process, which we expect is widely applicable in astronomy. Second, we aim to produce an estimate for the rate of FRBs that combines data across several surveys. Section \ref{nhpp} describes the general NHPP model and the specifics of our approach for FRB surveys. Section \ref{results} describes the results of the fit to the FRB survey data. Finally, Section \ref{discussion} presents some discussion on the model and what can be added in a more sophisticated analysis.

\section{The Nonhomogeneous Poisson Process Model} \label{nhpp}

A homogeneous Poisson process is a counting process that says that (1) the number of events that occur in an interval (e.g. of time or space or some other domain) follows a Poisson distribution whose mean is proportional to the size of that interval (e.g. length in 1-D, area in 2-D, etc.) and that (2) the number of events in two disjoint intervals are independent of each other. Thus, for a region of the sky, a telescope's field of view, for example, with area $A$, we might think that the number of FRBs occurring in this region over an interval of time $T$ is described by the Poisson distribution:
\begin{equation}
P\{ N(A,T) = n \} = \frac{1}{n!}(\lambda A T)^n e^{- \lambda A T},
\end{equation}
where the parameter $\lambda$ is interpreted as the rate for infinitesimally small area and length of time. The mean in this case, $\lambda A T$ is just the integral of the constant rate over the spatial and temporal intervals.

The nonhomogeneous Poisson model assumes that the instantaneous rate may vary over the space of observations. For example, in practice, a radio telescope has varying sensitivity over it's field of view. Assuming a radially symmetric sensitivity, the constant rate in the above example will need to be replaced by $\lambda(r)$, which varies with the radius from the center of the beam. The $\lambda A T$ term would be replaced by an integral over the desired region. We will flesh out the detail below when we consider a rate that varies with both the radius and the flux measurement itself. For more details on Poisson processes, homogeneous and otherwise, see \cite{kingman1993poisson} or \cite{ross1996stochastic}.

We assume that observations are made at a mix of single-dish radio telescopes (e.g. Parkes) and radio interferometers (e.g. VLA). We begin with a distinction between incident and observed flux. The incident flux is the flux that arrives at the telescope and that would be measured if the event were detected at the center of the beam. Let $\Lambda_I(s_I)$ be the cumulative rate of events with incident flux, that is the flux independent of attenuation from the observing device, greater than $s_I$. Assuming a uniform spatial distribution of possible events, we know that this should follow a power law with an exponent of $-3/2$. We will consider a general power-law form:
\begin{equation} 
\Lambda_I(s_I) = a s_{I}^{-b}
\end{equation}
where $b = 3/2$ is the so-called Euclidean scaling that is valid for any population with a fixed luminosity distribution, as long as it has a uniform spatial distribution (Vedantham et al 2016).
Let $\lambda_I(s_I)$ be the instantaneous rate for events with incident flux $s_I$ which is given by the derivative of $\Lambda_I(0) - \Lambda_I(s)$ with respect to $s_I$:
\begin{equation}
\lambda_I(s_I) = a b s_I^{-b-1} ds_I
\end{equation}

Our actual measurements are also affected by the telescope or array that we use. For simplicity, we will assume that the beam attenuation is radially symmetric and given by a function $p(r)$. The flux we measure, $s_O$, at some radius $r$ from the center of the beam is the product of the incident flux and and the attenuation function $s_O = s_I p(r)$. We can make substitutions to determine the instantaneous rate for observed flux of value $s_O$ at radius $r$:
\begin{equation}
\lambda_O(s_O, r) = ab \left( \frac{s_O}{p(r)} \right)^{-b-1} \frac{1}{p(r)} 2 \pi r ds_O dr dt.
\end{equation}
The $\frac{1}{p(r)} ds_O$ term replaces the $ds_I$ term as part of the change of variables and $2 \pi r dr$ is the infinitesimal observing area of the beam at radius $r$. The $dt$ acknowledges that the process has a time dimension, although the rate is constant in time.
For simplicity, we will assume a Gaussian beam shape for all telescopes, $p(r) = 2^{-(r/R)^2}$, where $R$ is the half-max radius for the beam.
This gives a rate function of
\begin{equation} \label{rate}
\lambda_O(s_O, r) = ab \left( s_O 2^{(r/R)^2} \right)^{-b-1} 2^{(r/R)^2} 2 \pi r
\end{equation}
For the beams of a single-dish telescope, the radius is not observed. For these devices, we need to integrate this function over the radius to produce the relevant rate function:
\begin{equation} \label{eq:ratesd}
\lambda_{O,SD}(s_O) = \frac{\pi R^2}{b \mbox{ln}(2)} ab s_O^{-b-1}.
\end{equation}
The $\frac{\pi R^2}{b \ln(2)}$ term can be interpreted as the effective area of the receiver.

We will take a maximum likelihood approach to estimating the model. In brief, we will write down the probability for the model, plug in the observed data, and then maximize this equation over the unknown parameters. The probability of observing $n$ events with fluxes $s_1, \cdots, s_n$ over $T$ hours at a single-dish telescope with minimum detectible flux $\sigma$ is described by
\begin{equation}
P\{N=n, \vec  S = (s_1, \cdots, s_n)\} \propto \exp \left\{ - T \Lambda_O(\sigma) \right\} \prod_{j=1}^{n} \lambda_{O,SD}(s_j),
\end{equation}
where $\Lambda_O(\sigma) = \int_{\sigma}^{\infty} \lambda_{O,SD}(s_O) ds_O$ is the cumulative rate for observed fluxes greater than $\sigma$.
There are two useful intuitive ways to derive this probability. The first is to consider the Poisson probability of the number events that will occur in the observing time and above the flux limit (the exponential term) and then, conditional on this number, consider the probability that these events occur at the observed fluxes (the product of $\lambda$ term). The second approach is to consider dividing the flux line into tiny intervals and treating $\lambda_{O,SD}(s_O)$ as constant over each interval. Now, each interval has a homogenous Poisson process over an interval small enough that the probability of more than one event is negligible. Letting the size of these intervals go to zero gives the above probability. For interferometers, the probability includes the measured radially location for each FRB:
\begin{equation}
P\{N=n, \vec  S = (s_1, \cdots, s_n), \vec R = (r_1, \cdots, r_n)\} \propto \exp \left\{ - T \Lambda_O(\sigma) \right\} \prod_{j=1}^{n} \lambda_O(s_j, r_j).
\end{equation}

Analysis is done on the log scale and the important quantity is the log-likelihood. Consider the log-likelihood component for survey $i$ with flux $n_i$ flux measurements $\vec s$:
\begin{equation} \label{eq:sdloglik}
\ell_i(a, b; \vec s) = -T_i \int_{\sigma_i}^{\infty} \lambda_{O, SD}(u; a, b) du + \sum_{j=1}^{n_i} \ln \lambda_{O, SD}(s_j; a, b)
\end{equation}
for single-dish observations and
\begin{equation} \label{eq:intloglik}
\ell_i(a, b; \vec s, \vec r) = -T_i \int_{\sigma_i}^{\infty} \int_{0}^{\infty} \lambda_{O}(u, v; a, b) dv du + \sum_{j=1}^{n_i} \ln \lambda_{O}(s_j, r_j; a, b)
\end{equation}
for interferometer observations with radius measurements $\vec r$. The log-likelihood over all surveys is simply the sum of these:
\begin{equation}
\ell(a, b; s, r) = \sum_i \ell_i(a, b; s, r).
\end{equation}
This function is maximized over the unknown parameters to obtain estimates for $a$ and $b$. This can be done with gradient-based numerical optimization scheme.
The covariance of these parameters is estimated using the second derivative matrix of this function: $V(\hat a, \hat b) = \left[ - \ddot \ell (\hat a, \hat b) \right]^{-1}$. See Theorem 5.1 of \cite{lehmann1998} for details of the asymptotic normality of these estimates and the rest of the book for background on the theory of maximum likelihood. Details of these computations are provided in the appendix.

\section{Results for FRB Surveys} \label{results}
Our data are given in Table \ref{tab:data}. The data come from 12 surveys using four telescopes and recording a total of 15 FRBs. This list does not include all surveys or detections, but only those for which we could determine the relevant observing characteristics. The list does include a variety of surveys, including some with zero detections. It also includes nearly 90\% of the published FRB detections. The table lists the number of FRBs, the observed fluxes, the minimum detectable flux, the time-on-sky, and the full width at half maximum for each instrument. The time-on-sky includes the number of beams for the single-dish instruments (for example, if the Parkes telescope observed for $x$ hours, the time listed here would be $13x$ to account for the the 13 beams). For simplicity, the beams are assumed to not overlap. We have focused the work on flux thresholds instead of fluence. Where necessary we follow \cite{law2015millisecond} and \cite{burke2016limits} by assuming a pulse width of 3 ms based on the average value for the FRBs detected to date.

\begin{table}[htp]
\begin{center}
\begin{tabular}{|c|c|c|c|c|c|}
\hline
Survey Ref. & n & flux & sensitivity & time & FWHM \\
 & & $s_i$ (Jy) & $\sigma_i$ (Jy) & $T_i$ (hrs) & $2  R_i$ (arc min) \\
\hline
\cite{lorimer2007bright} & 1 & 30 & 0.295 & 6376.4 & 14 \\
\hline
\cite{deneva2009arecibo} & 0 & NA & 0.1825 & 3213.7 & 4 \\
\hline
\cite{keane2010further} & 0 & NA & 0.2235 & 20247.5 & 14 \\
\hline
\cite{siemion2011allen} & 0 & NA & 59 & 4074 & 150 \\
\hline
\cite{burgay2012perseus} & 0 & NA & 0.2645 & 6923.8 & 14 \\
\hline
\cite{petroff2014absence} & 0 & NA & 0.434 & 12038 & 14 \\
\hline
\cite{spitler2014fast} & 1 & 0.4 & 0.105 & 80523.1 & 4 \\
\hline
\cite{burke2014galactic} & 1 & 0.3 & 0.3075 & 11924.9 & 14\\
\hline
\cite{ravi2015fast} & 1 & 2.1 & 0.2775 & 889.2 & 14 \\
\hline
\cite{petroff2015real} & 1 & 0.47 & 0.2775 & 1111.5 & 14 \\
\hline
\cite{law2015millisecond} & 0 & NA & 0.12 & 166 & 30 \\
\hline
\cite{champion2016five} & 10 & 1.3, 0.4, 0.5, 0.5, & 0.28 & 36224.5 & 14 \\
& & 0.87, 0.42, 0.69 & & & \\
& & 1.67, 0.18, 2.2 & & & \\
\hline
\end{tabular}
\end{center}
\caption{Survey data for several published FRB searches with 15 total detections. Except for the information for \cite{law2015millisecond}, the values for n, sensitivity, time are derived from \cite{burke2016limits}, which assumes a pulse width of 3 ms. Flux values are taken directly from each reference. The flux for \cite{ravi2015fast} was read from Figure 3 of that reference. The first four fluxes for \cite{champion2016five} are taken from \cite{thornton2013population}. The last five were read from Figure 1 of the reference. Note that the FRB Catalog includes two other detections, but the survey characteristics were difficult to obtain.}
\label{tab:data}
\end{table}

The observed data are plugged into the log-likelihood described above. Note that the only interferometer survey, \cite{law2015millisecond}, has no detections, so the instantaneous rate including a radius is not used here (the summation in Equation \ref{eq:intloglik} is not used). The likelihood is maximized at the values $\hat a = \ahat$ events per sky per day above a flux of 1 Jy and power-law index $\hat b = \bhat$. The 95\% confidence intervals are (271.86, 923.72) for $a$ and (0.57, 1.25) for $b$. The latter is notable as it excludes the Euclidean scaling value of 1.5.

These estimates give distinctly lower rates than other commonly cited results. \cite{champion2016five} gives a ``fluence-complete rate" of 2500 FRBs per sky per day above about 2 Jy-ms, or above 0.67 Jy assuming a 3 ms pulse. Their 95\% confidence interval is 900 to 5700 FRBs per sky per day. The corresponding rate from our model is only about 849.29 FRBs per sky per day with a 95\% confidence interval between about 508.86 and 1417.47 FRBs per sky per day. These intervals overlap, indicating some agreement. \cite{rane2016search} uses information from a collection of Parkes surveys and reports a rate of 4400 FRBs per sky per day above 4.0 Jy-ms, or 1.33 Jy assuming a 3ms pulse. They report a 99\% confidence interval of 1300 to 9600 FRBs per sky per day. Our corresponding estimate is about 451.51 FRBs per sky per day with a 99\% confidence interval between about 225.21 and 905.20 FRBs per sky per day. These intervals do not overlap suggesting a significant difference. These differences arise, at least in part, because our approach has included more surveys from a variety of telescopes, with more zero-detection results. Further, the methods are quite different with our approach attempting to leverage the information contained in the actual flux measurements.

\begin{figure}[h]
\begin{center}
\includegraphics{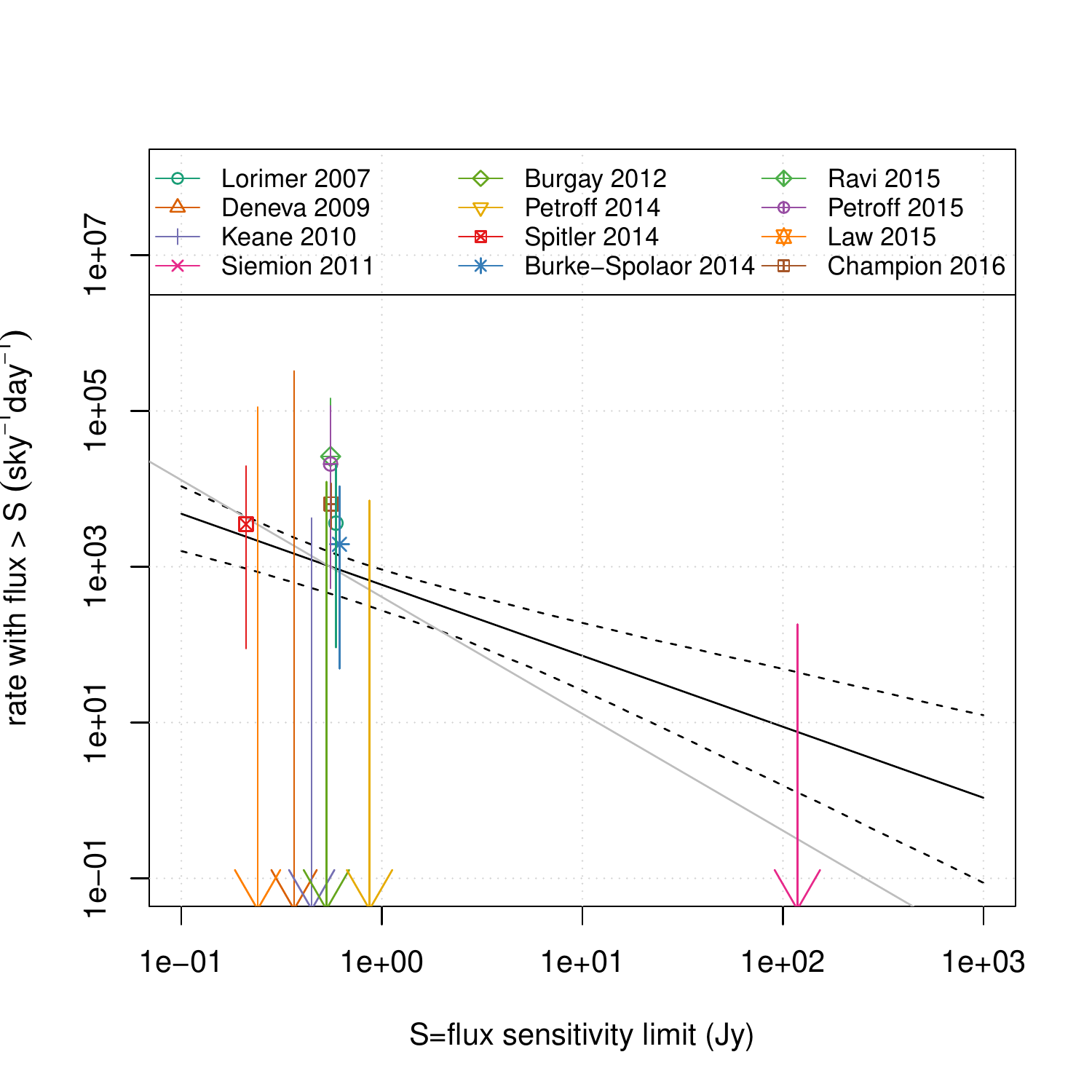}
\caption{The solid black line shows the fit from maximum likelihood estimator. The dashed black line shows the 95\% confidence bounds on the fit. The gray line is the best fit result obtained under the Euclidean scaling assumption (fixing $b=1.5$). The colored points show rate estimates from the individual surveys, with FOVs and sensitivities computed at the FWHM. and the associated bars show 95\% confidence bounds. Surveys with no detections provide only an upper bound and are indicated with arrows. The end points of the arrows are placed at the 97.5\% confidence value to match the other results.} 
\label{fig:fit}
\end{center}
\end{figure}

Figure \ref{fig:fit} shows the fit in terms of the cumulative rate as a function of incident flux, along with individual point estimates and uncertainties from each survey. The solid black line shows the maximum likelihood estimate and the dashed black lines show the 95\% confidence interval for the fit (see the appendix for details on the uncertainty of the cumulative rate function). The gray line shows the maximum likelihood fit under the constraint of Euclidean scaling (fix $b=1.5$). This line falls outside of our confidence bounds, again giving some indication that the data do not support Euclidean scaling.

The individual survey results are plotted at their FWHM sensitivity values, as is commonly done. Their rate values are computed as described in Section \ref{intro}, with the observing area assumed, as is commonly done, to be the area of the circle with diameter given by the full width at half maximum. This approach can be somewhat problematic however. Equation \ref{eq:ratesd} indicates that this area needs to be scaled by $b \ln(2)$. If $b$ takes the assumed value of $3/2$, the scaling term is about $1.04$ and the simple area calculation slightly overestimates the true effective area and estimated rates are not much effected. If we use the maximum likelihood estimate of $\hat b = \bhat$, the scaling term is $0.63$ and the simple area calculation underestimates the effective area by about $60\%$. This has the effect of overestimating the effective rate by about $60\%$.

Figure \ref{fig:fit} highlights the advantage of using the flux values. A crude estimate of the power-law model could be determined using the traditionally calculated individual rates by simply drawing a line through the various points. However, this approach is difficult, in part at least, because most of the surveys have similar flux thresholds. This makes it difficult to constrain the parameters, particularly the power-law index value. This would create the potential awkward desire to obtain detections using less sensitive surveys (those with higher thresholds) in order to constrain the index. The NHPP approach uses the flux values themselves to very directly estimate the power-law model.

To check the quality of the model fit, we will compare the fitted survival function with the empirical survival function. In this case, the survival function $P(s,b)$ gives the probability of observing an FRB with flux greater than $s$. As shown below, the function is dependent on $b$, but not $a$. No detections currently come from interferometers, but if some did, we recommend treating the position as unknown and using the observed flux with the single-dish rate function for this model check.

To derive this function, consider the Poisson process over the combination of all of the surveys. The rate function for this process is just the sum of the individual rate functions. Given that an FRB is detected in a survey, its flux value has a probability density proportional to the sum of the rate functions. The normalizing constant for this density is just the sum of the cumulative rate functions evaluated at the minimum sensitivity. To find the survival probability at a flux $s$, we just integrate each rate function from $s$ to $\infty$. Thus, survival function is given by
\begin{eqnarray} \label{eq:surv}
P(S > s; b) & = & \frac{\sum_i T_i \Lambda_{i,O}(\max[s, \sigma_i])}{\sum_i T_i \Lambda_{i,O}(\sigma_i)}, \\
                  & = & \frac{ \sum_i T_i \frac{\pi R_i^2}{b \ln(2)} \max[s, \sigma_i]^{-b}}{\sum_i T_i \frac{\pi R_i^2}{b \ln(2)} \sigma_i^{-b}}
\end{eqnarray}
where $i$ index surveys, $\Lambda_{i,O}$ is the cumulative rate function for observed flux for survey $i$, and $\sigma_i$ is the minimum observable flux for survey $i$. The $\max[s, \sigma_i]$ term arises because survey $i$ can only see FRBs with flux greater than $s$ when $s < \sigma_i$.

\begin{figure}[h]
\begin{center}
\includegraphics{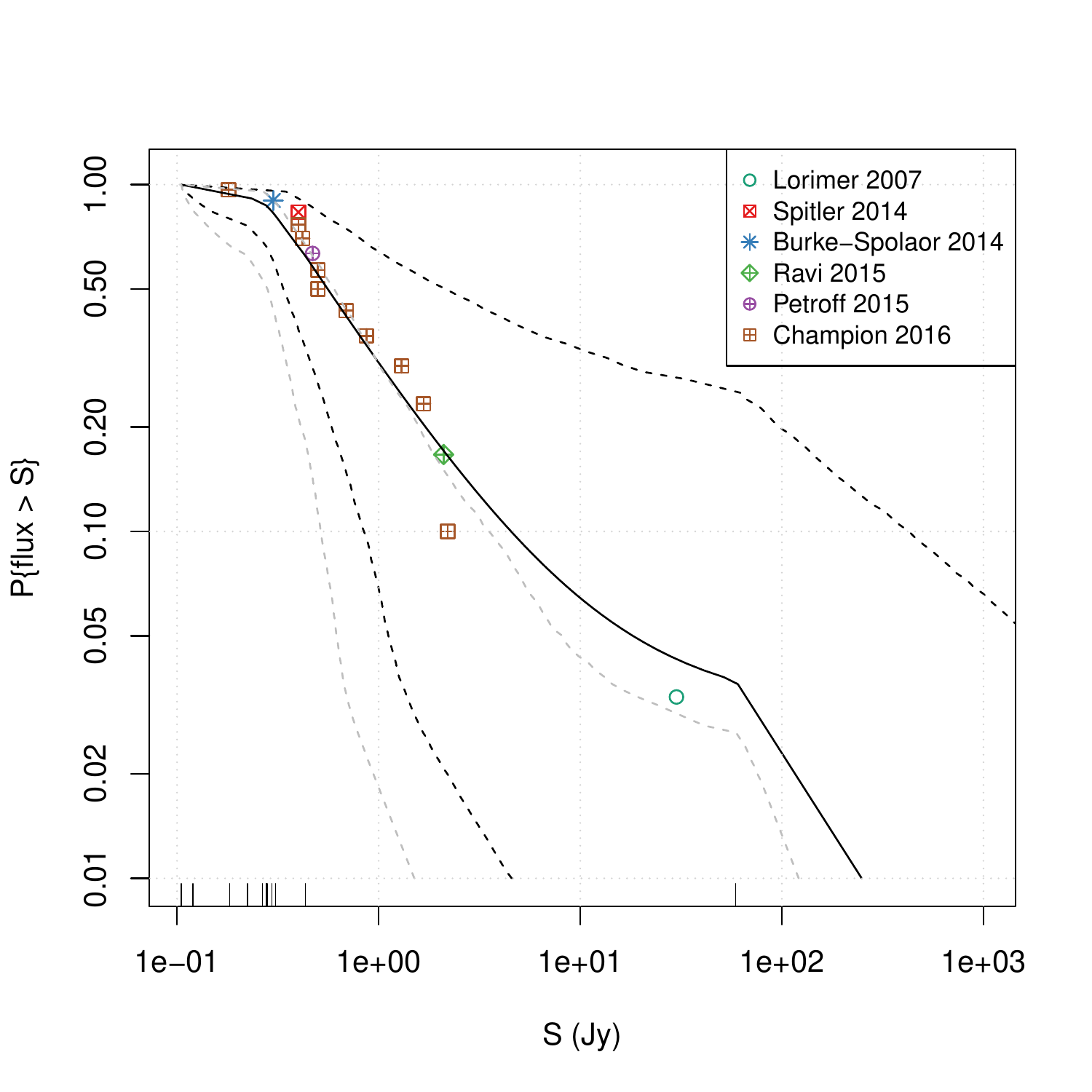}
\caption{Plot of the estimated survival function (solid black) with uncertainty (dashed black). The dashed gray lines indicate the uncertainty envelope based on assumption of Euclidean scaling. The individual points indicate observed FRBs. They are plotted at the observed fluxes on the $x$-axis. The $y$-axis shows the estimated empirical cumulative survivor function. The plot indicates no lack of fit to the power-law model. Dashes along the bottom axis indicate the sensitivity of each survey and correspond to locations where the survivor function suddenly becomes steeper because an additional survey term in Equation \ref{eq:surv} begins to decline with $s$.}
\label{fig:surv}
\end{center}
\end{figure}

Figure \ref{fig:surv} shows the survival function for the estimated model. The $x$-axis is flux and the $y$-axis is the probability of an FRB with greater flux. 
The solid black line shows the maximum likelihood estimate and the dashed black lines show the 95\% confidence interval from a parametric bootstrap (see the appendix). Let $s_{(i)}$ denote the $i$-th ordered observed flux. This is a non-parametric estimate of the $(i-0.5)/N$ flux quantile, for $N$ total observed fluxes.
Thus, the $y$-axis value for the $i$-th point sorted by flux is $(N - i + 0.5)/N$. This plot is useful as a diagnostic tool for model checking as it compares quantiles from the fitted model with the empirical quantiles. As all of the points fall close to the solid line and well within the confidence interval, there doesn't seem to be a lack of fit for the power-law model. The gray dashed lines shows the 95\% confidence interval based on the Euclidean scaling assumption. As before, the plot indicates that the data do not support this model.

The data in Table \ref{tab:data} and software used to make the FRB rate estimate will be available online soon.

\section{Discussion} \label{discussion}
The NHPP model provides an approach that can use all of the available information from astronomical events by giving probabilities for both the number of events detected in a survey and the measured flux values of those events. This is especially important when events, like FRB detections, are infrequently observed across a number of surveys; the model provides a natural solution to combining all of the available data across multiple single-dish and interferometer surveys. This approach avoids the need to make conservative choices about sensitivity (e.g. computing the sensitivity and field-of-view for the FWHM) that throw away information. Using the NHPP model, we estimate a rate of $\ahat$ sky$^{-1}$ day$^{-1}$ above a flux of 1 Jy. This rate is significantly lower than described in other work for two reasons. First, other rate estimates have not considered how non-Euclidean luminosity distributions can change the effective volume probed. In the best-fit model discussed here, this increases the effective volume by 60\%, which reduces the rate by a similar factor. Second, this rate estimate includes surveys from all points on the sky and does not attempt to model Galactic latitude dependence (discussed below).

One important use for this model is survey planning. We can estimate the expected number of hours until a detection using Equation \ref{eq:ratesd}. In the case of the VLA, this equation still applies assuming a uniform distribution for the event over the beam. We integrate this equation over $s_O$ to get the cumulative rate
$$
\Lambda_{O}(\sigma) = \frac{\pi R^2}{b \mbox{ln}(2)} a \sigma^{-b},
$$
and plug in the minimum sensitivity value to get a Poisson rate for a particular telescope. We can incorporate all sources of uncertainty by sampling $a$ and $b$ using the parametric bootstrap to get rate and then drawing an exponential random variable using the resulting rate. Assuming the sensitivity and FWHM values used for \cite{spitler2014fast}, \cite{petroff2015real}, and \cite{law2015millisecond}, we can get estimates until the time-to-first detection for Arecibo, Parkes, and the VLA. The mean time-to-first detection for these instruments is about 44880, 8459, and 861 hours respectively. In the case of Parkes, this value is reasonably close to the total number of hours divided by the total number of detections, 6838 hours per event, across all surveys in our data set. The 95\% upper bounds on the time-to-first detection are 140117, 26104, and 2693 hours for Arecibo, Parkes, and the VLA respectively.

The method presented here can be extended to include anything that systematically varies the sensitivity of a detection including dispersion measure scalloping, unknown pulse widths, or galactic position. The method can also be extended to include differences in sensitivities and effective areas from different pointings within the same survey. The approach can also be extended to a slightly generalized probability model. The negative binomial model is also a counting model, which can be derived from the Poisson model by letting the rate follow a Gamma distribution. This introduces extra variation that can be used to account for effects that are not included in the rate model.

An important extension could be the inclusion of more precise beam shapes. We have assumed radially symmetric Gaussian beams, that do not overlap in the case of the single-dish telescopes. The approach can be extended to include numerical integration of precise descriptions of the beam, including two-dimensional descriptions of the multi-beam single dish telecopes that properly include beam overlap. When we have integrated over unknown event locations, we have done so by integrating the radial position from zero to infinity. Both the shape and the limits of integration are primarily for convenience, but we believe that the error from these approximations, as tested by by truncating the Gaussian and considered smaller effective FWHM, is very small. 

A second important extension is the inclusion of dependence on galactic latitude. We can crudely consider latitude dependence by estimating the model parameters using just the surveys that collected data exclusively or mostly at high galactic latitudes. We include \cite{lorimer2007bright}, \cite{ravi2015fast}, \cite{petroff2015real}, \cite{law2015millisecond}, and \cite{champion2016five} in this group. These surveys produce an estimate of $\hat a=1314.96$ FRBs per sky per day above 1 Jy (95\% CI: 509.83, 2152.37) and a power-law index of $\hat b =0.79$ (95\% CI: 0.38, 1.22). The 1 Jy rate rate is higher and the index is lower. Our predicted rate above 0.67 Jy is about 1814.18 FRBs per sky per day with a 95\% confidence interval between about 1029.68 and 3196.40, which falls entirely within the interval given by \cite{champion2016five} at this flux. Our predicted rate above 1.33 Jy is about 1046.52 FRBs per sky per day with a 99\% confidence interval between about 500.80 and 2186.89, which does now overlap with the interval given by \cite{rane2016search}, but their rate is computed using data over a range of latitudes and their rate is given as an all-sky rate. While this suggests some dependence, we are not comfortable stating this with certainty without building latitude into the model more directly.

This model also finds evidence for a non-Euclidean flux distribution of FRBs with a power-law index of $\bhat$. This is consistent with estimates made with other techniques and data \citep{vedantham2016associating}. A non-Euclidean distribution produces unintuitive properties for observed FRB properties, such as that it is easier to detect more distant events and that small, wide-field radio telescopes can detect more events than large, sensitive telescopes. Finally, we note that a flat flux distribution is inherently surprising, as demonstrated by the extreme brightness of the first FRB reported by \cite{lorimer2007bright}.


\section*{Acknowledgements}
This work was supported by the University of California Lab Fees Program under award number \#LF-12-237863.

\appendix

\section{Gradient and Hessian of the Log-Likelihood}

Here we derive the gradient and Hessian for the log-likelihood. The latter is important for computing an estimate of the variance of the unknown parameters. For convenience, we will reparameterize the model in terms of $\alpha = ab$ and $\beta = b+1$.

Recall that the log-likelihood is the sum of the log-likelihoods from each survey:
\begin{equation}
\ell(\alpha, \beta; s, r) = \sum_i \ell_i(\alpha, \beta; s, r),
\end{equation}
thus, we can consider the derivatives of each summand individually. Consider the gradient of one survey-level log-likelihood, a vector of the partial derivatives.
\begin{equation}
\dot \ell_i(\alpha, \beta; s, r) = - T_i \int_{\sigma_i}^{\infty} \int_{0}^{\infty} \dot \lambda_O(u,v; \alpha, \beta)dv du +  \sum_{j=1}^{n_i} \frac{\dot \lambda_{O}(s_j, r_j; \alpha, \beta)}{\lambda_{O}(s_j, r_j; \alpha, \beta)}
\end{equation}
The result for a single-dish term is similar. The key thing to note is that this function only depends on the derivatives of the rate function. The same is true of the second derivative matrix, the Hessian, which only depends on the first and second derivatives of the rate function,
\begin{equation}
\ddot \ell_i(\alpha, \beta; s, r) = - T_i \int_{\sigma_i}^{\infty} \int_{0}^{\infty} \ddot \lambda_O(u,v; \alpha, \beta)dv du + \sum_{j=1}^{n_i} \left[ \frac{\ddot \lambda_{O}(s_j, r_j; \alpha, \beta)}{\lambda_{O}(s_j, r_j; \alpha, \beta)} -   \frac{\dot \lambda_{O}(s_j, r_j; \alpha, \beta) \dot \lambda_{O}(s_j, r_j; \alpha, \beta)^{T}}{\lambda^2_{O}(s_j, r_j; \alpha, \beta)} \right].
\end{equation}
Note that $\dot \ell_i(\alpha, \beta; s, r)$ is column vector of length two and that $\ddot \ell_i(\alpha, \beta; s, r)$ is a $2 \times 2$ matrix.

Here are the first and second partial derivative terms for $\lambda_{O,SD}$
\begin{eqnarray}
\frac{\partial}{\partial \alpha} & = & \frac{\pi R^2}{\ln(2)} \frac{s_O^{-\beta}}{\beta-1}\\
\frac{\partial}{\partial \beta} & = & - \frac{\pi R^2}{\ln(2)} \frac{\alpha s_O^{-\beta}}{\beta-1} \left[ \ln(s) + \frac{1}{\beta-1}\right] \\
\frac{\partial^2}{\partial \alpha^2} & = & 0 \\
\frac{\partial^2}{\partial \alpha \partial \beta} & = & - \frac{\pi R^2}{\ln(2)} \frac{s_O^{-\beta}}{\beta-1} \left[ \ln(s) + \frac{1}{\beta-1}\right] \\
\frac{\partial^2}{\partial \beta^2} & = & \frac{\pi R^2}{\ln(2)} \frac{\alpha s_O^{-\beta}}{\beta-1} \left[ \ln^2(s_O) + \frac{2\ln(s_O)}{\beta-1} + \frac{2}{(\beta-1)^2} \right]
\end{eqnarray}

Here are the first and second partial derivative terms for $\lambda_{O}$
\begin{eqnarray}
\frac{\partial}{\partial \alpha} & = & 2^{(r/R)^2 + 1} \pi r \left( s 2^{(r/R)^2} \right)^{-\beta} \\
\frac{\partial}{\partial \beta} & = & - 2^{(r/R)^2 + 1} \pi r \alpha \left( s 2^{(r/R)^2} \right)^{-\beta} \left[ \ln(s) + \left(\frac{r}{R}\right)^2 \ln(2) \right] \\
\frac{\partial^2}{\partial \alpha^2} & = & 0 \\
\frac{\partial^2}{\partial \alpha \partial \beta} & = & - 2^{(r/R)^2 + 1} \pi r \left( s 2^{(r/R)^2} \right)^{-\beta} \left[ \ln(s) + \left(\frac{r}{R}\right)^2 \ln(2) \right]\\
\frac{\partial^2}{\partial \beta^2} & = & 2^{(r/R)^2 + 1} \pi r \left( s 2^{(r/R)^2} \right)^{-\beta} \left[ \ln(s) + \left(\frac{r}{R}\right)^2 \ln(2) \right]^2
\end{eqnarray}

The gradient is helpful for maximizing the likelihood. The Hessian is necessary for the covariance of the parameters, as described above.

\section{Confidence Intervals Based on the Parametric Bootstrap}

Figures \ref{fig:fit} and \ref{fig:surv} show confidence intervals for the cumulative rate of incident flux and the survival plot. Both sets of bounds are computed using the parametric bootstrap. This uses the covariance of the estimated parameters give in Section \ref{nhpp}. The statistical theory says that our unknown parameters are approximately normally distributed with mean given by the maximum likelihood estimates and covariance given by $V(\hat \alpha, \hat \beta) = \left[ - \ddot \ell (\hat \alpha, \hat \beta) \right]^{-1}$. The parametric bootstrap propagates this uncertainty by sampling the parameters from this distribution and computing a quantity of interest (e.g. the cumulative rate over some grid of flux values) for each sample. This provides an empirical distribution for the quantity of interest from which we can compute quantiles to use as bounds.

\bibliographystyle{apalike}
\bibliography{frb}

\end{document}